\documentclass{PoS}

\usepackage{url}
\graphicspath{{figures/}}

\newcommand{\be}{\begin{equation}}
\newcommand{\ee}{\end{equation}}
\newcommand{\bea}{\begin{eqnarray}}
\newcommand{\eea}{\end{eqnarray}}
\newcommand{\non}{\nonumber}

\title{String tension at finite temperature Lattice QCD}

\ShortTitle{String tension at finite temperature Lattice QCD}

\author{ \speaker{Pedro Bicudo} \\
CFTP, Instituto Superior T\'ecnico, Universidade T\'ecnica de Lisboa \\
E-mail: \email{bicudo@ist.utl.pt}
}

\author{Nuno Cardoso \\
CFTP, Instituto Superior T\'ecnico, Universidade T\'ecnica de Lisboa \\
E-mail: \email{nunocardoso@cftp.ist.utl.pt}
}
\author{Orlando Oliveira \\
CFC, Universidade de Coimbra \\
E-mail: \email{orlando@teor.fis.uc.pt}
}

\author{Paulo J. Silva \\
CFC, Universidade de Coimbra \\
E-mail: \email{psilva@teor.fis.uc.pt}
}

\abstract{
The critical curve of string tension as a function of the temperature is computed in SU(3) Lattice QCD.
We present the results for the string tension utilizing a pair of Polyakov loop and antiloop, with two different techniques.
We compare the colour averaged loop-antiloop which is gauge invariant but is only adequate to study the string tension, and the colour singlet loop-antiloop using the Landau gauge fixing which also enables to compute the coulomb part of the free energy.
}

\FullConference{ The XXIX International Symposium on Lattice Field Theory - Lattice 2011\\
July 10-16, 2011\\
Squaw Valley, Lake Tahoe, California}

\begin{document}

\section{Introduction and Motivation}

\begin{figure}
     \begin{center}
       \includegraphics[width=0.7\columnwidth]{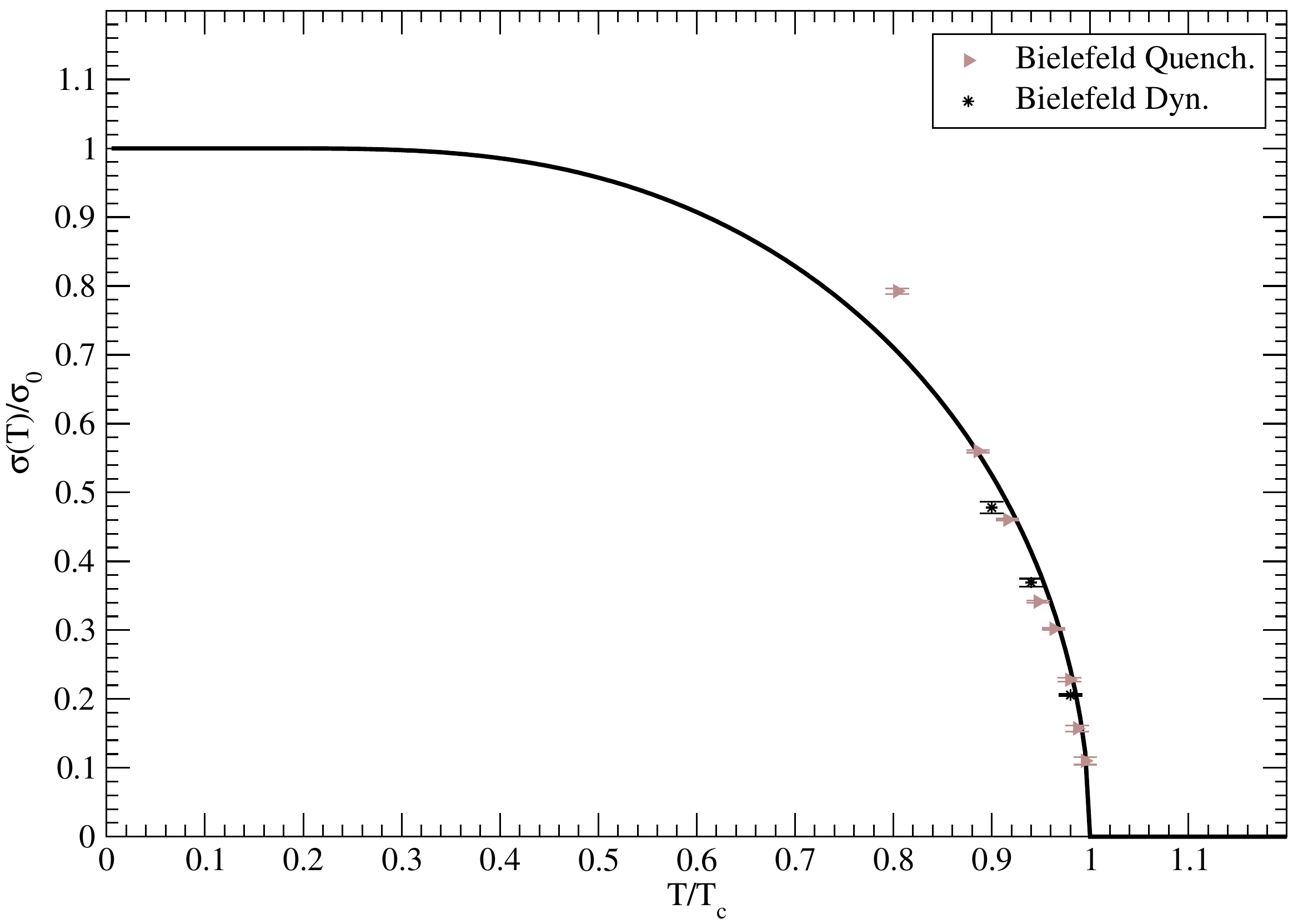}
     \end{center}
\caption{$\sigma(T)$ Bielefeld data and emprirical curve}
\label{sigma_AL}
\end{figure}

The confinement/deconfinement transition and chiral symmetry restoration are non-perturbative phenomena, 
elevant for heavy ion collisions and for quark models at finite temperature.

The nature of the confinement/deconfinement phase transition has been investigated in Lattice QCD using as order parameter the Polyakov Loop. For quenched QCD, the lattice data points towards first transition. However the single Polyakov loop is more adequate for detailed studies of  temperatures $T>T_c$.

Here we study the string tension $\sigma$, adequate to study temperatures $T<T_c$. In the formulation of quark models  
\cite{Bicudo:2011cb}, 
the { \em value of the string tension $\sigma$}  determines the linear confining potential, with $\sigma \ne 0$ in the confined, low temperature phase and $\sigma = 0$ above the critical temperature $T_c$ (deconfined phase). 
Further, { \em chiral symmetry} can also be related with $\sigma$
\cite{Bicudo:2011qs}.

Therefore, the
string tension is also an
order parameter for the
deconfinement and chiral phase transition, which can be
measured via two Polyakov loops. 

The Bielefeld lattice QCD group measured the quark-antiquark free energy in a wide range 
of temperatures $T > 0.8 T_c$ and also measured the critical curve 
$\sigma (T)$
\cite{Kaczmarek:1999mm,Doring:2007uh}.
In Fig. \ref{sigma_AL} we depict the Bielefeld results,
together with the empirical fit of Bicudo 
\cite{Bicudo:2010hg}.

In order to
understand the nature of the deconfinement phase transition,
here we aim to
measure $\sigma (T) $ also for temperatures smaller than $0.8 T_c$.

To generate our configurations and to compute the Polyakov loops,
we utilize GPUs of the NVIDIA FERMI generation
\cite{Cardoso:2010di, Cardoso:2011}
and the TFlops cluster Milipeia
\cite{downloads}.

\begin{figure}
\begin{center}
       \includegraphics[width=0.7\columnwidth]{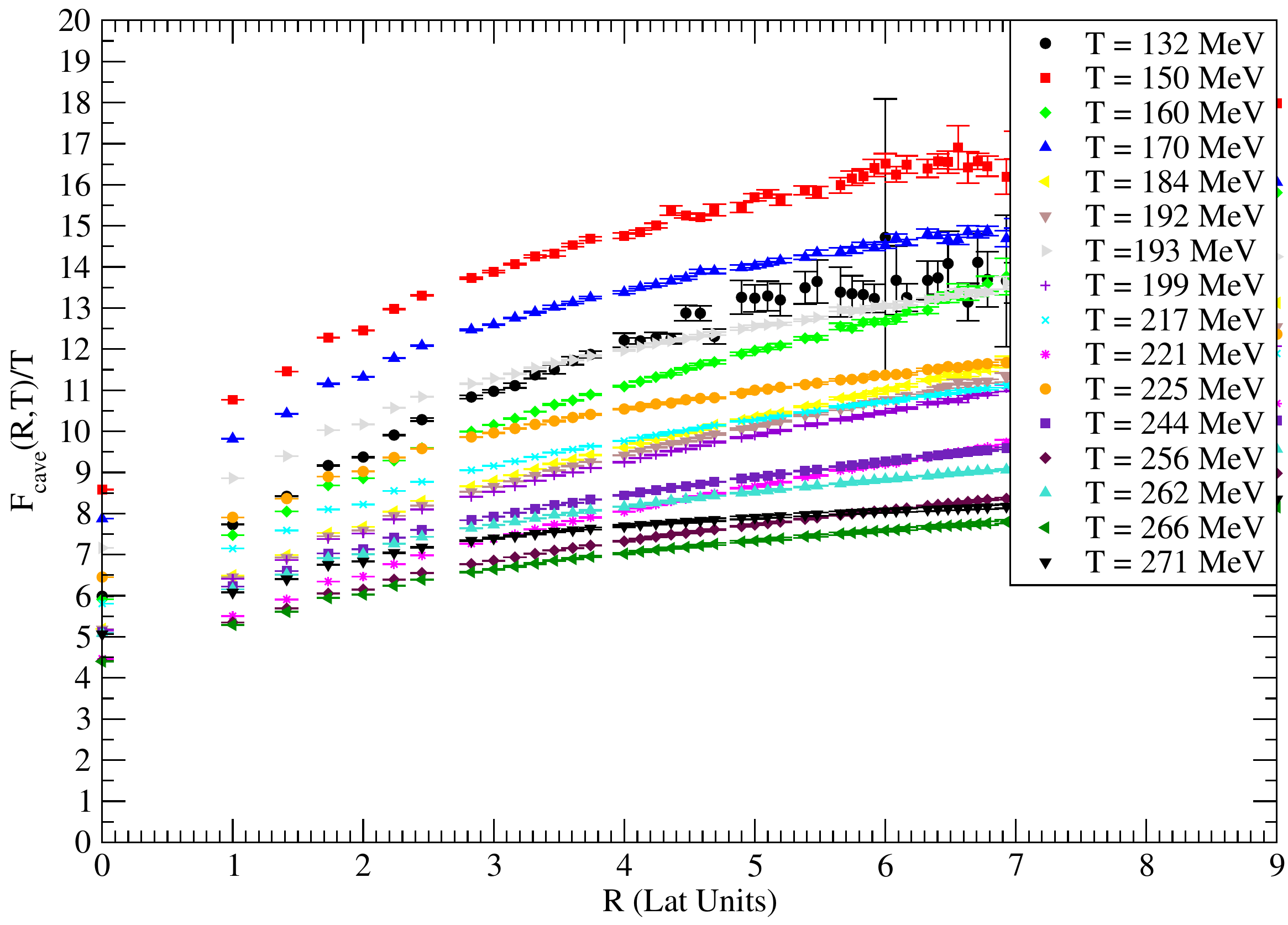}
     \end{center}
\caption{Color average free energy with the link integration method}
\label{Fcave_T_Lisboa}
\end{figure}

\begin{figure}
\begin{center}
       \includegraphics[width=0.7\columnwidth]{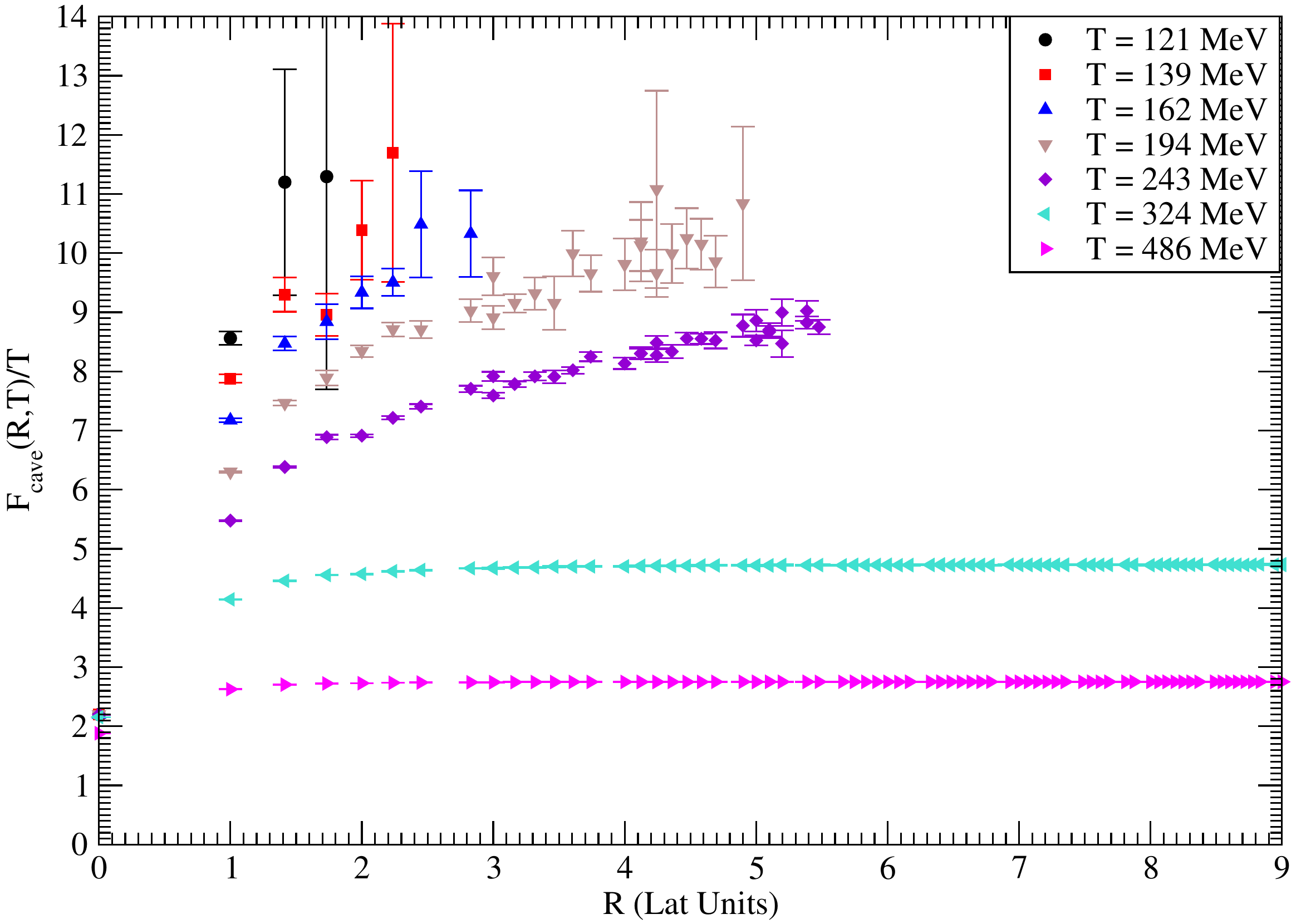}
     \end{center}
\caption{Color average free energy with the Landau gauge fixing}
\label{Fcave_T_Coimbra}
\end{figure}

\begin{figure}
     \begin{center}
       \includegraphics[width=0.7\columnwidth]{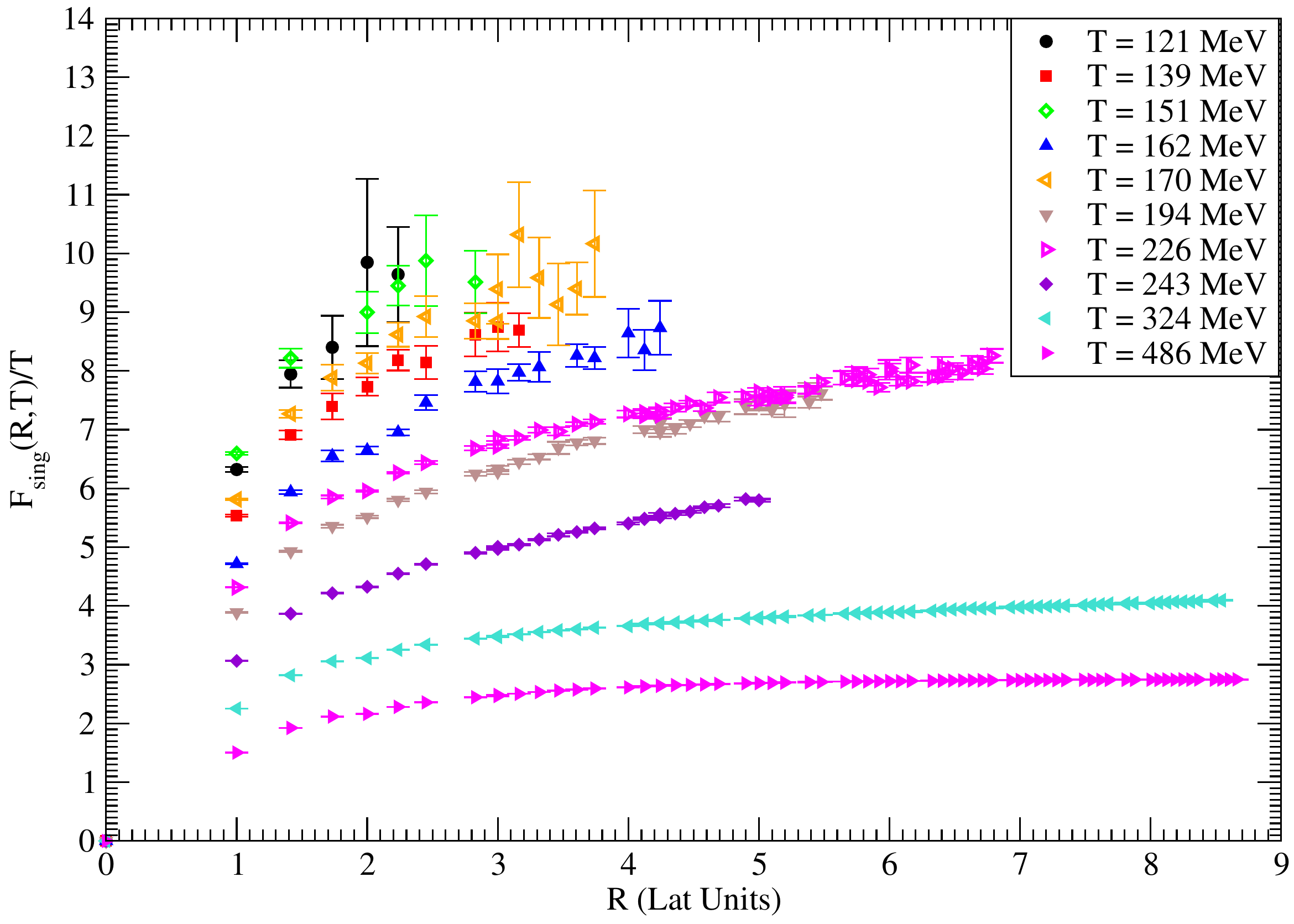}
     \end{center}
\caption{Color singlet free energy with the Landau gauge fixing}
\label{Fsing_T_Coimbra}
\end{figure}

\begin{table}[b!]
\centering
\begin{tabular}{lll}
\hline
\textbf{Polyakov Loop}                  &
    $\qquad L(\vec{r}) $ & =  $ ~ U_t(\vec{r},0) \, U_t(\vec{r},1) \, \cdots \, U_t(\vec{r},N_t-1) $   \\
                 &
   &   $\propto  ~ \exp\left( - F_q / T\right)$   \\
\textbf{Singlet Correlator}              &
   $\qquad Re \, \left\{ Tr \left [\langle L(\vec{0}) L^\dagger (\vec{r}) \rangle \right] \right\}$
                                                         & $\propto ~ \exp\left( - F_1(\vec{r},T) / T\right)$ \\
\textbf{Color Average Correlator}  &
   $\qquad \langle Re \, \left\{ Tr \left [ L(\vec{0}) \right] \right\} ~ ~
                                                    Re \, \left\{ Tr \left [ L^\dagger (\vec{r}) \rangle \right] \right\} $
                                                         & $\propto ~ \exp\left( - F_{ave} (\vec{r},T) / T\right)$ \\
\textbf{Singlet Free Energy}         & $\qquad F_1 (\vec{r}, T) $ & \\
\textbf{Octet Free Energy}  & $\qquad F_8 (\vec{r}, T) $ & \\
\textbf{Color Average Free Energy}  & $\qquad F_{ave} (\vec{r}, T) $ & = $~ \frac{1}{9} F_1 + \frac{8}{9} F_8$\\
\hline
\end{tabular}
\caption{Color average and color singlet free energies.}
\label{tab:def}
\end{table}

\vspace{2cm}
\section{Framework}

We extract the string tension form the static quark-antiquark free energy. The free energy is computed with a pair of Polyakov loop-antiloop. 
We explore two different color approaches to the Polyakov loop-antiloop, defined in Table \ref{tab:def}.

The color average Polyakov loop-antiloop pair, leading to the color average free energy is  gauge invariant, but at shorter distances the free energy may be contaminated by the color octet contribution. 

On the other hand, the Singlet Free Energy is the one we aim at, however it is composed by the product of two non gauge invariant Wilson line,
and in order to evaluate it we have to fix the gauge, choosing a particular gauge. Here we use the Landau gauge. 

In this poster  we apply both the gauge invariant calculations color average technique and we also utilize the Landau gauge technique.

\begin{figure}
     \begin{center}
       \includegraphics[width=0.7\columnwidth]{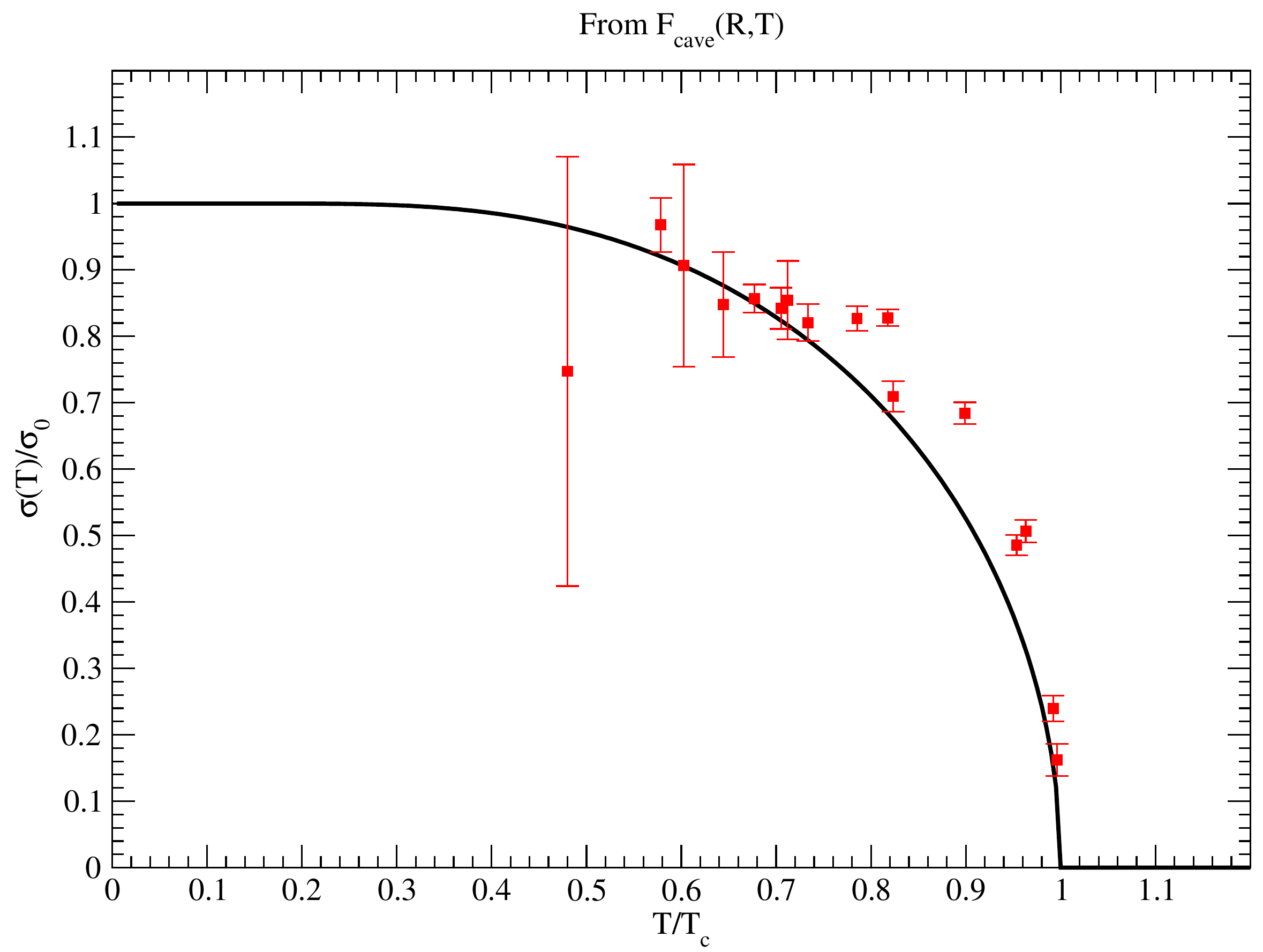}
     \end{center}
\caption{$\sigma(T)$ for $T \in 0.48 T_c - T_c$ extracted with the color average free energy}
\label{sigma_lisboa}
\end{figure}

We use the equation of the lattice spacing times the string tension at zero temperature, given by
\be
\left(a\sqrt\sigma\right)(g) = f(g^2)\left[ 1 + 0.2731 \,\hat{a}(g)^2 - 0.01545 \,\hat{a}(g)^4 + 0.01975 \,\hat{a}(g)^6 \right]/0.01364 
\ee
in Edwards et al. \cite{Edwards:1997xf} 
valid in the region $5.6 \leq \beta \leq 6.5$,
\bea
\hat{a}(g) &=& \frac{f(g^2)}{f(g^2(\beta=6.0))},
\non \\
f(g^2) &=& \left(b_0g^2\right)^{-\frac{b_1}{2b_0^2}} \exp\left( -\frac{1}{2b_0g^2}  \right),
\non \\
b_0 &=& \frac{11}{(4\pi)^2},\quad\quad b_1 = \frac{102}{(4\pi)^4}
\eea

Therefore, $T/T_c$ is given by
\be
\frac{T}{T_c}=\frac{\left(a\sqrt\sigma\right)(g(\beta_c))}{\left(a\sqrt\sigma\right)(g(\beta))}
\ee
where $\beta_c$ was obtained in Lucini et al.
\cite{Lucini:2003zr}.

The critical temperature 
\cite{Fodor:2001pe,Brower:1981vt} 
$T_c$, in the quenched case is circa 270 MeV, for 2  flavour dynamical fermions 170 MeV and for 3 flavour dynamical fermions150 MeV.
Here we study the critical curve $\sigma(T)$ in quenched SU(3) QCD.

\vspace{2cm}
\section{Results and conclusion}

\begin{figure}
     \begin{center}
       \includegraphics[width=0.7\columnwidth]{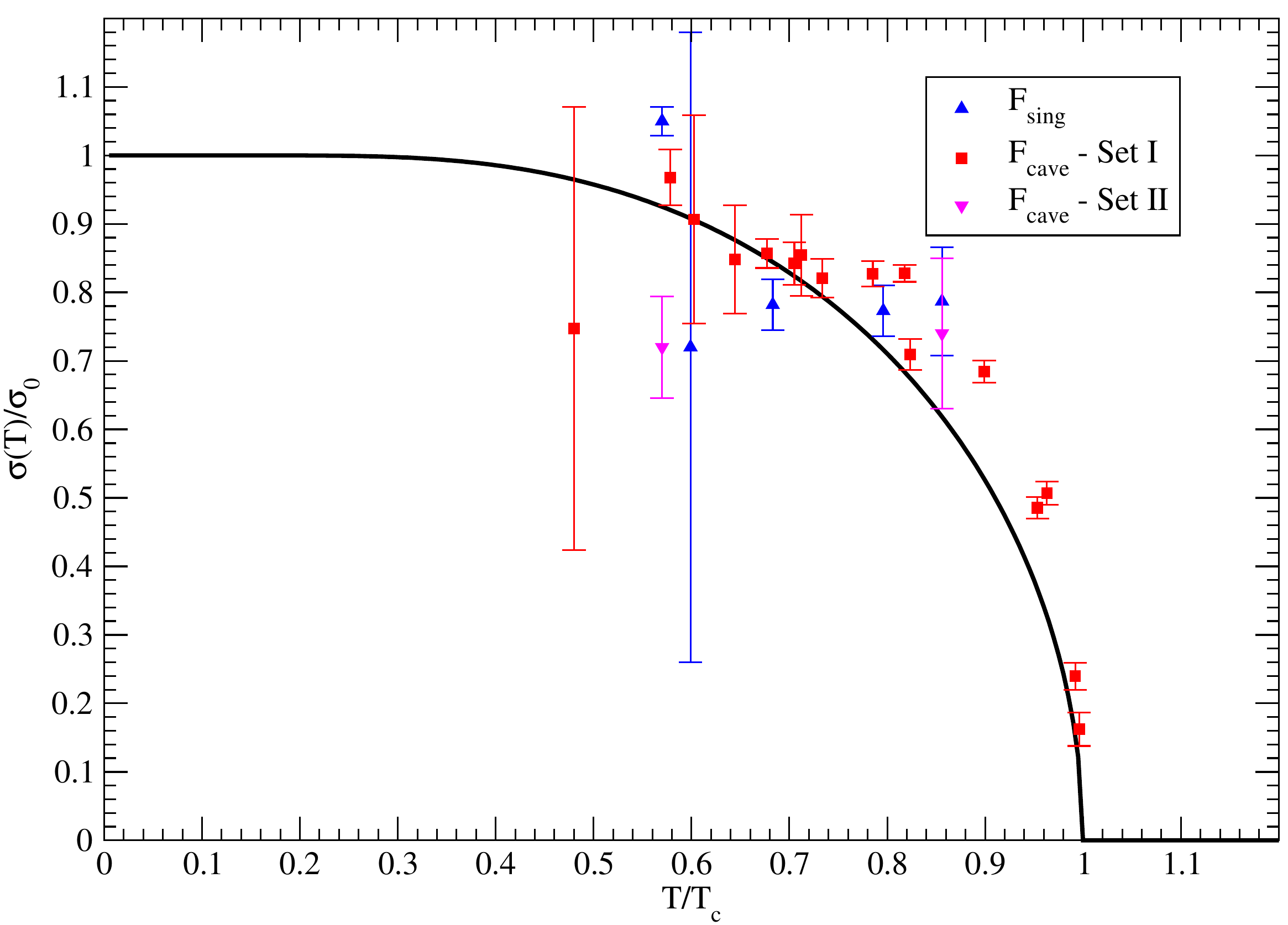}
     \end{center}
\caption{$\sigma(T)$ for $T \in 0.48 T_c - T_c$ extracted both with the color average free energy and with the color singlet free energy}
\label{sigma_all}
\end{figure}

\begin{figure}
     \begin{center}
      \ \ \ \ \ \ \includegraphics[width=0.83\columnwidth]{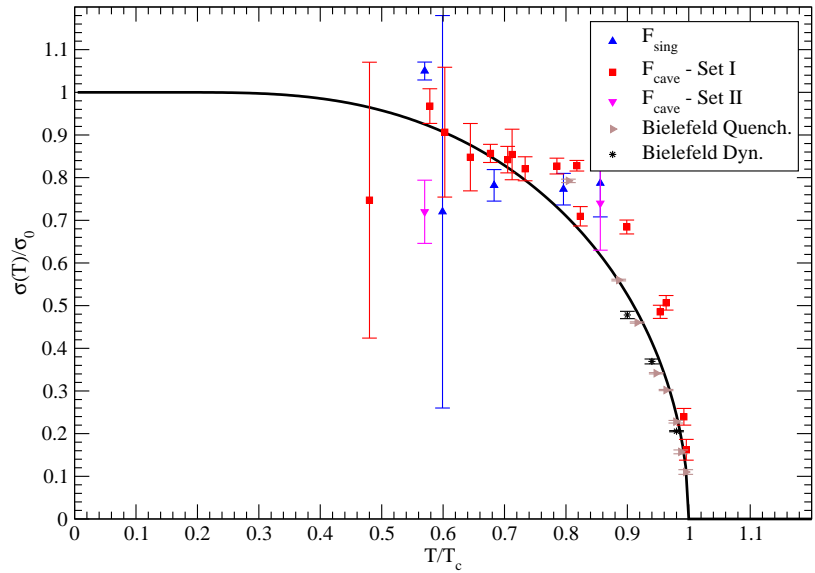}
     \end{center}
\caption{Summary of ours and Bielefeld results for $\sigma(T)$ at $T \in 0.48 T_c - T_c$ }
\label{sigma_PT_AL}
\end{figure}

 Our results are preliminary results only, aimed at exploring the string tension critical curve $\sigma(T)$ for finite temperatures below $T_c$,
in particular for the so far unexplored $T < 0.8 T_c$. 
We illustrate the different free energies in Figs \ref{Fcave_T_Lisboa},  \ref{Fcave_T_Coimbra} and  \ref{Fsing_T_Coimbra}.

The noise leads to quite large error bars in the string tension points of Figs. \ref{sigma_lisboa}, \ref{sigma_all} and \ref{sigma_PT_AL}.
We need to improve our signal to noise ratio, and to improve the precision of our temperatures, in order to improve the study of the critical curve.

Nevertheless, our preliminary results show that it is possible to study the critical curve of $\sigma(T)$ in a wider temperature range,
and so far the empirical critical curve continues to show good agreement with the expected values.

\acknowledgments
This work was partly funded by the FCT contracts, POCI/FP/81933/2007,
CERN/FP/83582/2008, PTDC/FIS/100968/2008, CERN/FP/109327/2009
and CERN/FP/116383/2010.
Nuno Cardoso is also supported by FCT under the contract SFRH/BD/44416/2008.
Paulo Silva is also supported by FCT under the contract SFRH/BPD/40998/2007.

\end{document}